# Reducing Spreadsheet Risk with *F*ormula*D*ata***S*leuth**®

Bill Bekenn and Ray Hooper
Fairway Associates Ltd. PO Box 846, Ipswich IP5 3TT, UK
E mail: info@fairwayassociates.co.uk  Web: www.fairwayassociates.co.uk

**ABSTRACT**

*A new MS Excel application has been developed which seeks to reduce the risks associated with the development, operation and auditing of Excel spreadsheets. FormulaDataSleuth® provides a means of checking spreadsheet formulas and data as they are developed or used, enabling the users to identify actual or potential errors quickly and thereby halt their propagation. In this paper, we will describe, with examples, how the application works and how it can be applied to reduce the risks associated with Excel spreadsheets.*

## 1. INTRODUCTION.

MS Excel, like many powerful and flexible computer applications is prone to "misuse" and error to which even experienced users are not immune. It has been estimated that 94% of spreadsheets have at least one error in them [Panko, 2005]. There are many cases of spreadsheet error opening up business and other organisations to serious financial risk [EuSpRIG, 2007] and [Croll, 2005] or worse [Croll, 2006]. Reductions in risk can be brought about by improved development practice [O'Beirne, 2005] and [Read & Batson, 1999] and training. Many Excel users are perhaps not fully aware of the vulnerabilities of the application and are thus likely to remain unaware of the need for training. The role of the spreadsheet auditor will grow as companies become increasingly concerned about their vulnerability to spreadsheet error. Developers working with rapidly changing requirements have a need for an application to assist and aid their work to prevent errors at source.

In this paper, *F*ormula*D*ata***S*leuth**® will be described. This is a new application produced to reduce the risks associated with the development, auditing and operation of MS Excel spreadsheets. Some of Excel's vulnerabilities will be described, including its methods of auditing, formula analysis and error detection. An overview of the major features of *F*ormula*D*ata***S*leuth**®, including those central to the reduction of spreadsheet risk, will then be presented. The principles of operation will be described including the WatchFormula sheet, which plays a pivotal role in the operation of the application. A simple example from telecommunications systems cost analysis will then be presented.

## 2. EXCEL VULNERABILITIES.

Excel has some features to help developers and auditors avoid or detect errors. However, in practice, these can be somewhat "hit and miss". Excel's security and protection features tend to be unused when spreadsheets are continually being changed. The "green triangles" that warn of formula irregularities do not work in every situation. For example, a range vulnerability occurs when adding rows, so that on inserting rows above a cell with a column summation of formulas, no green warning triangle appears to show that the range is missing the new row. The order in which cells are inserted and formulas are filled can also give rise to errors that are not flagged by a green triangle. The formula display feature, which places the formulas in the cells (as opposed to calculated values), can present the auditor with a daunting "spot the error" task in all but the simplest spreadsheets.

Structural issues can arise from the many different ways cells are referenced (relative, absolute, mixed or Name). Spreadsheet development and restructuring can be compromised by vulnerable reference dollaring causing errors in "copy and paste" and fill operations. The problems grow with the number of sheets and workbooks and inter-workbook links are a significant additional source of errors.

## 3. OVERVIEW OF *F*ormula*D*ata*S*leuth®.

*F*ormula*D*ata*S*leuth® is an Excel application in the form of a workbook that "watches" a target Excel spreadsheet, workbook or group of workbooks. Information from contiguous areas of formulas and data (a single cell is a special case of an area) is recorded. For formula areas (data areas are treated slightly differently), this information falls into one of three categories as shown in figure 1:

1. Area Elements - information about the elements of the formulas of those areas e.g. cell references,
2. Area Properties - information relating to the properties of the areas e.g. location,
3. Flags e.g. Error Found.

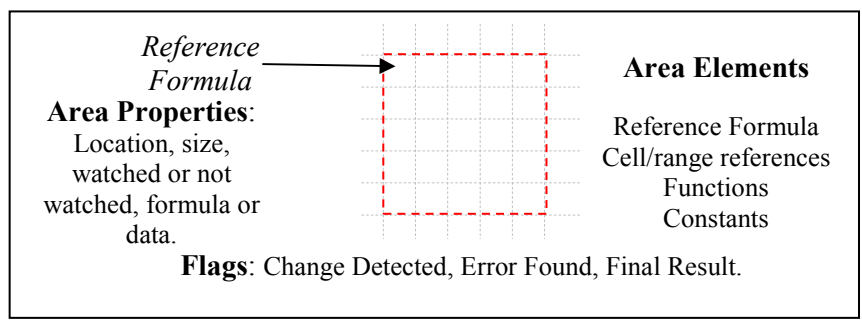

Figure 1. *F*ormula*D*ata*S*leuth® Elements and Properties of Contiguous Formula Areas.

Formula Area Elements are the components and terms of the formula. Note that Area Properties include whether the area is watched or not. Only watched areas are recorded, other areas are "Not Watched" by default. The user determines what is watched.

The information depicted in figure 1 together with seven linked procedures, described in section 4, is accessed through a special Sleuth toolbar (shown in section 5). The Area Elements, Properties and Flags are contained in the WatchFormula sheet that is added to the target workbook by the application.

The user interface is designed to allow "one-click" access to the application's procedures. Messages within user forms assist the user in choosing options and the course of action on flagging an error or change.

The application may be used as a development tool for new spreadsheets or as an improvement tool for existing ones. It can be used as an auditing tool and generates an audit trail. It may also be used in spreadsheet operation (when only data is being changed) to ensure users do not inadvertently damage a spreadsheet.

# 4. PRINCIPLES OF OPERATION.

The WatchFormula sheet is at the heart of *F*ormula*D*ata*S*leuth®. Information in the form of Elements, Properties and Flags is manipulated in the WatchFormula sheet by the action of procedures selected from the special Sleuth toolbar. These procedures are Error Checking and Reconciliation, Spreadsheet Restructuring and Updating, Multiple Insert/Delete, and, Sheet and Block Replication. These and other procedures will be illustrated in section 5.

## 4.1. The WatchFormula Sheet

This sheet records, in the target workbook, all the specific information about the Formula and Data Areas. For formula areas this information is:-

1. Area Elements Information
    a. the current formula string in R1C1 format for the area,
    b. the last watched formula string in R1C1 format for the area,
    c. the parsing of formulas into references (single cells or ranges and Excel Names),
    d. the details of inter-workbook links for destination references.
2. Area Properties Information
    a. the target sheet name,
    b. the location of the area/cell in terms of its extent, recorded as top left and bottom right rows and columns,
    c. the details of Excel Names referring to watched areas,
    d. the details of the sources of inter-workbook links,
    e. the name of any area that has been designated as a group (see section 4.4),
    f. the status, particularly whether it is a "final result" with no dependencies.
3. Flags
    a. the occurrence of changes to the Formula Areas,
    b. the occurrence of errors.

Areas of cells where a formula has been filled/pasted, so that all the cells contain the same generic formula (the reference formula of figure 1), are treated as a single formula. Selected Formula Areas appear in the WatchFormula sheet after "watching" is initiated from the Sleuth toolbar. Changes to watched areas are automatically recorded as they are made, but a record of the previous Formula Area is logged. It is possible to cancel the "watching" of any previously watched area from the Sleuth toolbar.

A summary of the WatchFormula sheet is shown in a CheckFormula Sheet entered in the target workbook automatically by the Sleuth. This sheet lists the formulas and data areas, facilitates rapid navigation and highlights "warnings" of irregularities as well as errors. In Excel 2003 the WatchFormula and CheckFormula sheets can handle up to 10,000 formula and data areas and 60 single cell or range references per formula area.

## 4.2 Error Checking and Reconciliation

The Error Found Flag is set True whenever the comparison of the current and previous watched areas shows irregularities in the Area Elements. Error checking is instigated from the Sleuth toolbar. There are three main types of formula irregularities that the procedure identifies:

1. Formulas that have been "damaged" or overwritten with data or another formula,

2. Formulas containing references to cells or ranges which have not been watched or don't exactly match a watched area; these are invalid precedents,
3. Data and Formulas in Areas, which are either not referred to by other watched formulas or referred to inconsistently; these are invalid dependents. Final Results are watched and have no dependents and are thus valid and not flagged as errors.

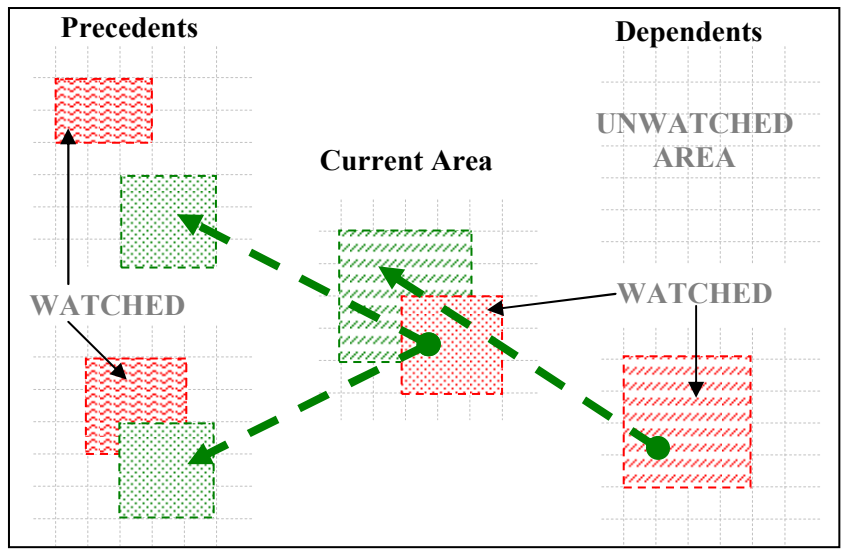

Figure 2. Reconciliation: Examples of Invalid Precedents and Dependents

In the case of damaged formulas only the Formula Elements of the Current and Watched Areas are compared. Figure 2 shows the more complex situation of reconciliation of precedents and dependents. The current watched area's (centre, red box) Precedents (left, green boxes) are examined. If the Area Properties of these Precedents are found to not exactly match those in the WatchFormula sheet then an error is flagged. Two examples of this situation are shown to the left in figure 2 where the properties (location, size etc.) of the precedent areas (left, green boxes) do not match with the watched areas (left, red boxes). This depicts precedents that are not watched (top) and partially watched (bottom) and are thus invalid.

The situation of the Dependents is similar to the Precedents with the current watched area (centre, green box), now being the Precedent, having its Dependent (right, red box) examined. Dependents can only be discovered by working backwards to find Precedents for all watched areas. There are four possible conditions:

1. there are, in fact, no dependents and,
    a. if the current area is a Formula Area it may be a Final Result,
    b. if the current area is a Data Area it is an error,
2. there are dependents but they are not watched and therefore it is not recorded that they exist (top right of figure 2),
3. there are watched dependents but they don't match exactly to the current watched area (bottom right of figure 2),
4. there are watched dependents that are matched and there is no obvious error.

Vulnerable reference dollaring is also flagged as an error or warning by the Sleuth. Examples of this include checking that a single cell reference is absolute and that single row/column references use absolute row/column.

## 4.3 Spreadsheet Restructuring and Updating

Spreadsheets are often extended or enhanced with a) additional rows or columns and b) the movement of formulas and data blocks within sheets and to different sheets. As additional data and formula areas are created they can be Watched using the Watch Formula (and Fill), Watch Data button. When blocks of formulas and data are moved (including between sheets) the WatchFormula sheet tracks the moves and re-calculates the current formulas for the moved areas and any dependent areas. The Check Watched Formulas button can be used at any time to see whether the movement of blocks (by "cut and paste" or "drag and drop") has caused any errors in the formulas. In most cases the Fix Selected Area/Cell, Accept Changes button can be used to correct any errors caused. However, where key cells have been deleted the Watch information may be damaged. In this case, the error will still be detected and can be corrected manually and then re-Watched. When areas of a sheet are to be completely re-worked the Watches can be deleted using the Delete (or clean up) Watch Entries button. Once the re-work is complete the Find Formulas button can find and highlight the new formulas, which can also be watched individually or automatically.

## 4.4 Multiple Insert /Delete

The Sleuth has a procedure for inserting or deleting rows or columns in multiple connected watched areas (i.e. replicated structured blocks; a common feature of spreadsheets). The connected areas can be in a spreadsheet, workbook or group of workbooks. This uses an Area Property defining each of the connected areas as a Group. The Sleuth toolbar's Group button allows the user to define the connected areas and enter a common Group name, which is recorded on the WatchFormula sheet. When additional rows or columns are required in all the connected areas, the row above (or the column to the left of) the insertion is selected in one of the areas and the Sleuth toolbar's Insert Rows/Columns Below/Right (in Group) is clicked. This inserts rows below or (columns to the right) in all the connected areas with that Group name. If the insertion occurs in a formula area, then the formula is filled from the Reference Formula of that watched area. Multiple row or column inserts are supported. This procedure is also available in the Sleuth toolbar for deletes and for the more normal insert row above (or column to the left).

The Sleuth's multiple insert/delete procedure is configured to obviate Excel's range vulnerability. If, for example, the selected row is just above a column summation, when the Insert Rows/Columns Below/Right (in Group) button is clicked an extra blank row is added below the inserted row and ranges are adjusted to include it. This extra row is greyed and will be checked to ensure it remains blank. There is then no possibility that any further inserted row will be missed in a summation or other function relating to the altered range, provided the Sleuth toolbar is used. Manual row insertion now also works avoiding the range vulnerability.

## 4.5 Sheet and Block Replication

The Sleuth has a procedure for replicating a set of watched areas (which can be in different sheets or workbooks); along with the replication of the information in the WatchFormula sheet. This is similar to Excel's "copy and paste" function, but in an intelligent form with checking and adjustment of the replicated areas. The adjustment ensures the reference dollaring can be set up in a way that fills correctly even if the dollaring selected would not normally "copy and paste" correctly. All areas that are replicated together end up referencing each other (adjusted even if absolute). Any references to un-replicated areas are left unchanged (un-adjusted even if relative). Overall the procedure obviates many of Excel's "copy and paste" vulnerabilities, in particular removing the need to change the reference dollaring for "copy and paste". The effect of relative/absolute referencing is overridden according to what else is being replicated at the same time.

**4.6 Data Areas.**

The Sleuth watches Data Areas as well as Formula Areas and automatically recognises them. It will identify the following errors in a watched Data Area:

1. blanks; the user has the option of accepting these as zeros,
2. formulas; these should not exist in Data Areas,
3. text data in numerical areas and numerical data in text areas,
4. data values which are out of a bounded range, where this range may be set by the user or automatically calculated by the Sleuth.

The purpose of the bounded range feature is to identify "rogue" data, but the automatic calculation can be refined to reflect statistical distribution of the numerical data.

**5. TELECOMMUNICATIONS COST MODELS**

Telecommunications cost models are complex. They bring together aspects of networks such as geography, technology, services, capacity, management and interconnection. These are all interrelated and affect costs. There are many design trade-offs and spreadsheet models are commonly used. The models are usually in a state of flux as the various scenarios are examined. Ensuring the validity of a model as it is modified is a vital requirement. A section from a typical Telecommunications Cost Model showing how equipment is costed is now described. Volumes of two types of customer connections are considered. The ports per equipment card and the volume enable the number of cards to be calculated for each year. SUMPRODUCT() is then used to calculate cost for each year.

**5.1 Restructuring the Cost Model**

Figure 3 shows the cost model before and after development modifications have added an additional card type (rows 4 & 8) and dragged and dropped the Year 0 column out of main view (column G to J). The model is fully watched by the Sleuth but the modifications have been made without it. The move of Year 0 has corrupted the formulas in G6:J8, but there is no indication (green triangle) of error.

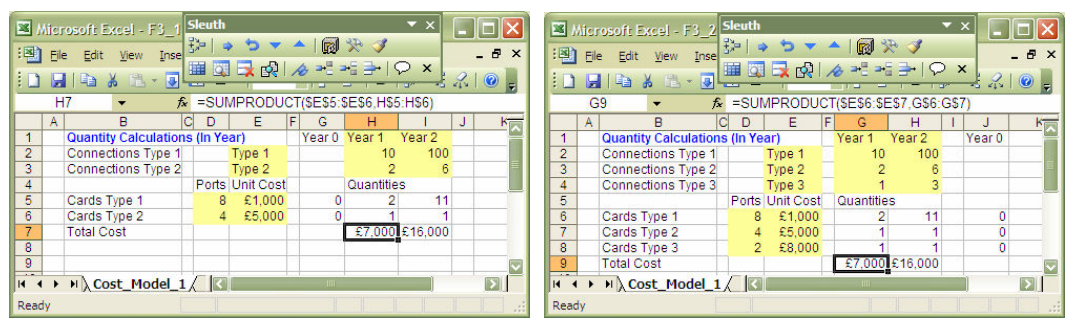

Figure 3. Equipment Model – Adding a Card & Moving Year 0

The formula bar shows that the SUMPRODUCT() function has not included the additional row, due to Excel's range vulnerability. Moreover, the area, G6:H8 will no longer fill right correctly:-

| Cell   | H5 moved to G6                        | I5 moved to H6                        |
|--------|---------------------------------------|---------------------------------------|
| Before | =ROUNDUP(H2/$D5,0)-SUM($G5:G5)        | =ROUNDUP(I2/$D5,0)-SUM($G5:H5)        |
| After  | =ROUNDUP(G2/$D6,0)-**SUM($J6:J6)**    | =ROUNDUP(H2/$D6,0)-**SUM($G6:G6)**    |
| Filled | =ROUNDUP(G2/$D6,0)-**SUM($J6:J6)**    | =ROUNDUP(H2/$D6,0)-**SUM($J6:K6)**    |

Clicking the Sleuth's "Check Watched Formulas" and "Go There" buttons reveals these errors (top screen of fig. 4), showing also that the new rows are not included in the watched areas. In Figure 4 the red borders and cross-hatching indicate errors and the orange borders show changes. These formatted indications on the target spreadsheet are temporary and can be removed. The errors listed are "<Error in Formula.>" because G6:H8 is no longer filled right, and "<Formula refers to a Cell/Area that is NOT Watched.>" because G6, if filled right, would then refer to K6 which is outside the watched Year 0.

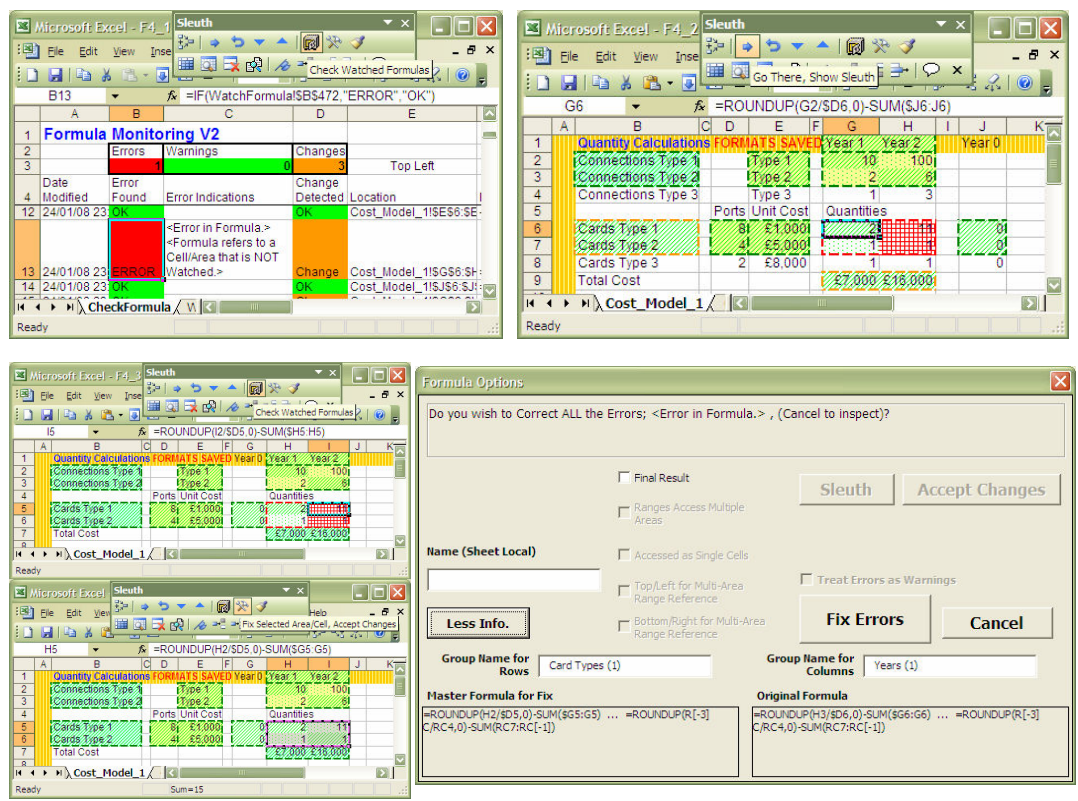

Figure 4. Sleuth Screenshot – Finding and Correcting Errors

Moving Year 0 back to column G and deleting the new rows restores the model to its original state, but there is now a residual error (I5:I6). This can be fixed by the Sleuth with the Fix Selected Area/Cell button. A form is displayed as in figure 4 and clicking "Fix Errors" removes the error in line with the information on the form.

An experienced Excel user might anticipate these problems and make the necessary adjustments. However, common spreadsheet manipulations such as these can be handled in a straightforward way by the Sleuth.

Instead of using the normal manual method in Excel, the Sleuth can insert the additional rows in a single operation by clicking the Insert Rows Below button (see figure 5). All formula areas are now automatically filled and the SUMPRODUCT() references are automatically adjusted. Additional blank rows are inserted to protect the spreadsheet from the range vulnerability. The original formatting of the spreadsheet is then restored using the Sleuth's "Restore Original Cell Formats" button.

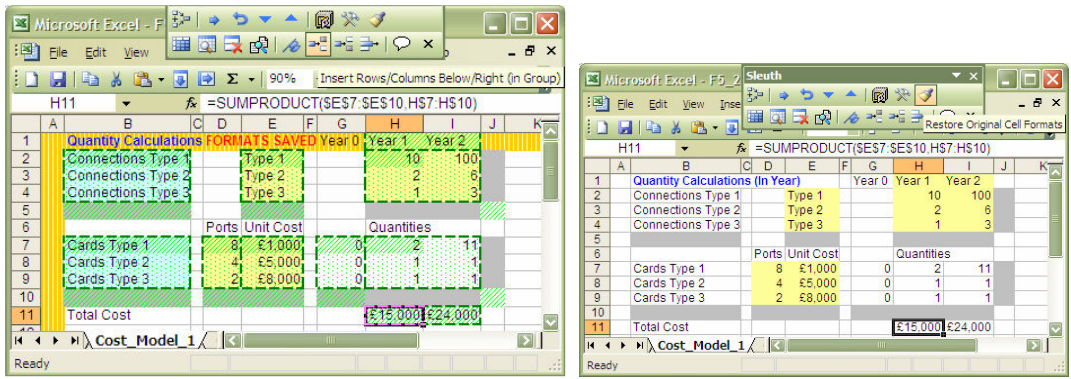

Figure 5. Inserting Rows and Columns with the Sleuth

### 5.2 Operating the Cost Model

The Sleuth has an Operational Mode, which has fewer features but still allows "Check Watched Formulas" to be run at any time and operational errors to be detected. If an operator, perhaps not aware of the vulnerability of formulas (even protected formulas) to cell movement, were to move the Year 2 data to the spare column J and update Year 2 (figure 6), then this will be detected by the Sleuth as shown in figure 7.

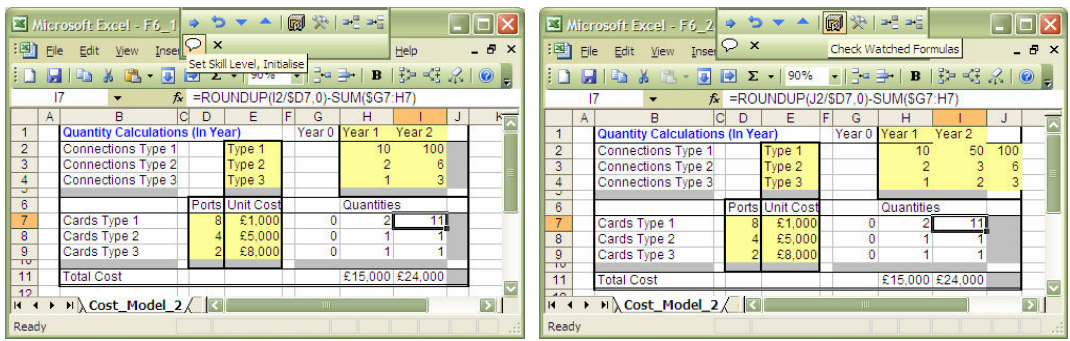

Figure 6. Moving the Year 2 Data

Clicking the "Check Watched Formulas" button finds the resulting errors "<Data entered over Blank.>" and "<Error in Formula.>" in the CheckFormula sheet.

The errors are fixed by alternately clicking "Go There" and "Fix Selected Area/Cell" buttons to produce the screens as indicated in figure 8. The first message box in figure 8 gives the information "<Data entered over Blank.>" error including the upper and lower bounds of the data set. The second message box provides the information "<Error in Formula.>" including the master formula for the fix.

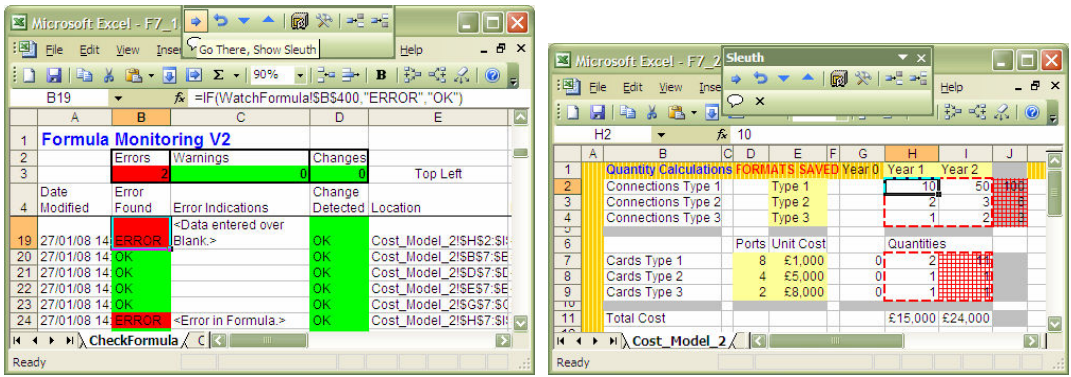

Figure 7. Errors Highlighted in the CheckFormula sheet and "Go There" (to the location).

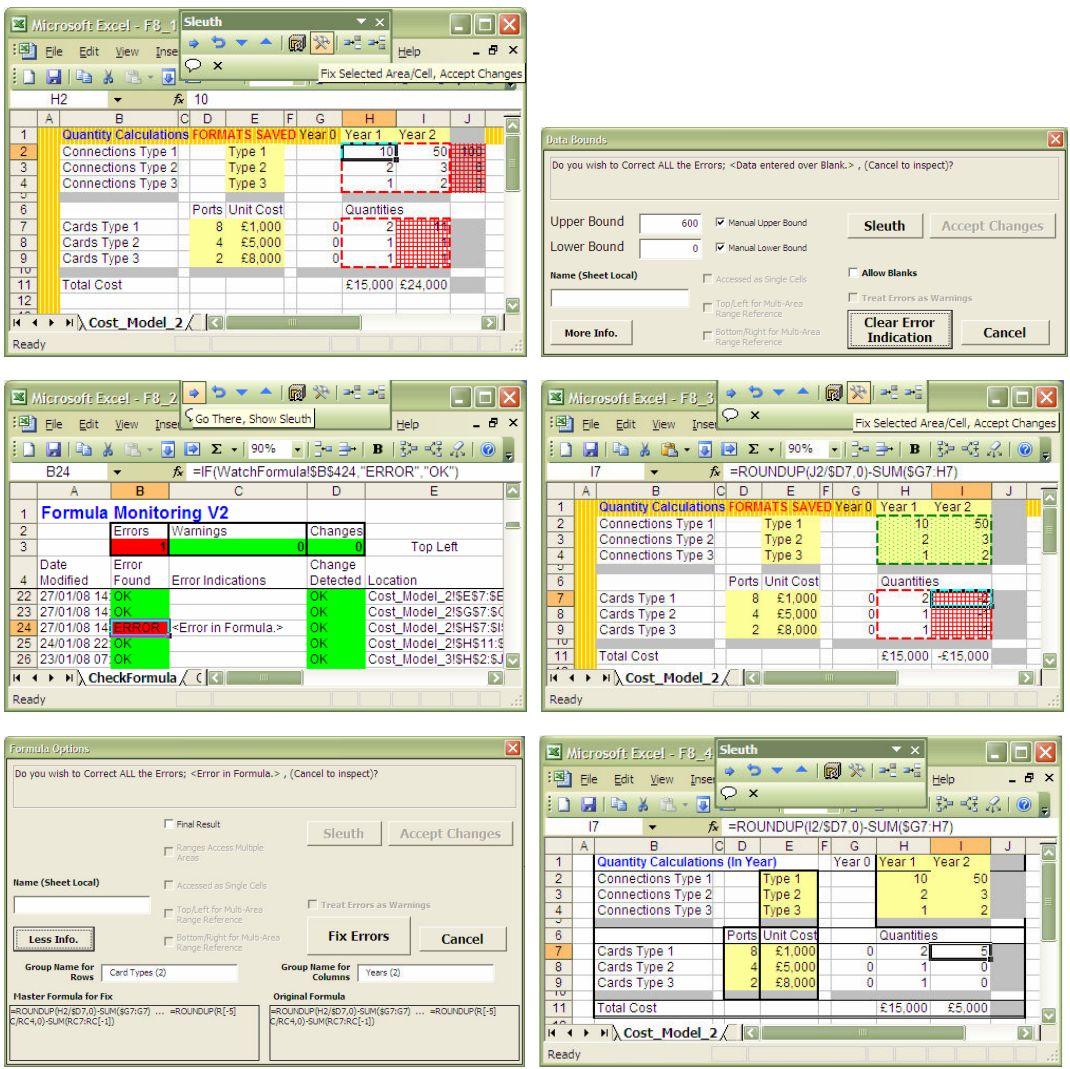

Figure 8. The "Go There", "Fix Selected Area/Cell" Repeating Sequence

In Operational mode the final "Fix Selected Area/Cell" operation automatically restores the sheet formatting.

### 5.3. Reconciliation and Detection of Errors

The model is now enhanced, in Development mode; separating cumulative and in-year calculations, adding a cost decline and calculating Shelves and Racks as well as Cards. Errors have been made in referencing the wrong cells and filling plus incorrect dollaring followed by copy and paste. The Sleuth has detected these as reconciliation errors and the results are shown in figure 9, including Excel auditing arrows to show referencing errors.

Figure 9. Cell Referencing Detected by the Sleuth – Excel Auditing Arrows also Shown

## 5.4 Tracing References

The model has now (with the help of the Sleuth during the restructuring) been split into two sheets separating the quantity and cost calculations. The following set of steps show how the "Trace References" and "Go There" procedures are used to analyse and trace the references from an answer back to the source data. The final step in figure 10 shows how the Trace References procedure breaks down a more complex formula. It evaluates the functions contained within it separately and shows how functions are nested.

Figure 10. The "Trace References" and In-Depth Formula Analysis

## 6. CONCLUSION

The Excel application *F*ormula*D*ata*S*leuth® has been described highlighting its procedures and features, which diminish or remove the susceptibility of spreadsheets to Excel's many vulnerabilities and consequent errors. These features include:

1. all encompassing error condition detection that includes,
    a. formula damage by overwrite,
    b. end to end reconciliation of precedents and dependents in formula references,
    c. sophisticated data error detection.
2. a simple user interface for,
    a. watching of all formula and data areas,
    b. full error checking at any time,
    c. intelligent error fixing,
    d. powerful auditing via detailed formula analysis
3. procedures which aid error-free spreadsheet development using,
    a. multiple insert/delete in related blocks,
    b. sheet and block replication complete with the Sleuth watch information.

In this way, many of the risks in spreadsheet development, auditing and operation are reduced. Examples from the world of Telecoms Cost Modelling have been examined but this type of model is very similar to others in the infrastructure-costing domain. Such models often undergo continual development and are almost never static as they seek to examine many different scenarios. Spreadsheet costing model results are extremely important as they can determine a business's investment in new technology, services and systems. Mistakes or errors are best prevented at source not least because there is little time for auditing and rework later.

The key to successful use of an application of this type is a simple and effective user interface. If risk is to be reduced without constraining users in their spreadsheet development and without having to check and audit extensively, there is a need for an application which will assist, aid and then generate trust. *F*ormula*D*ata*S*leuth® has been developed with these principles in mind.